\documentclass[12pt]{article}
\usepackage{amsmath}
\usepackage{amssymb}
\usepackage{array}
\usepackage{geometry}
\usepackage{graphicx}
\usepackage{enumerate}
\usepackage[small, compact] {titlesec}
\numberwithin{equation}{section}
\begin{document}
\vskip7cm\noindent
\begin{center}{\bf On a Generalized Entropy Measure Leading to the Pathway Model:\\ 
with a preliminary application to solar neutrino data}\\
\vskip.3cm{A.M. Mathai}\\
 \vskip.2cm{Centre for Mathematical and Statistical Sciences}\\
 \vskip.1cm{Peechi Campus, KFRI, Peechi 680653, Kerala, India}\\
mathai@math.mcgill.ca\\
 \vskip.2cm{and Department of Mathematics and Statistics, McGill University Canada}\\
 \vskip.1cm{805 Sherbrooke Street West, Montreal, Quebec, Canada, H3A2K6}\\
\vskip.2cm{and}\\
 \vskip.2cm{H.J. Haubold\\
 \vskip.1cm Office for Outer Space Affairs, United Nations,\\
 \vskip.1cm P.O. Box 500, Vienna International Center, A-1400 Vienna, Austria}\\
hans.haubold@gmail.com\\
 \vskip.2cm{and Centre for Mathematical and Statistical Sciences}\\
 \vskip.1cm{Peechi Campus, KFRI, Peechi 680653, Kerala, India}\\
\end{center}

\vskip.5cm\noindent{\bf Abstract} \vskip.3cm An entropy for the
scalar variable case, parallel to Havrda-Charvat entropy was
introduced by the first author and the properties and its connection
to Tsallis non-extensive statistical mechanics
and the Mathai pathway model were examined by the authors in previous papers. In the
current paper we extend the entropy to cover scalar case,
multivariable case, and matrix variate case. Then this measure is
optimized under different types of restrictions and a number
of models in the multivariable case and matrix variable case are
obtained. Connections of these models to problems in
statistical, physical, and engineering sciences are also pointed out. An application of 
the simplest case of the pathway model to the interpretation of solar neutrino data
is provided. 



\vskip.5cm\noindent{\bf 1.\hskip.3cm Introduction} \vskip.3cm
Classical Shannon entropy has been generalized in many directions [11]. An
$\alpha$-generalized entropy, parallel to Havrda-Charvat entropy,
introduced by the first author, is found to be quite useful in
deriving pathway models [6], including Tsallis statistics [10] and
superstatistics [1,2]. It is
also connected to Kerride's measure of inaccuracy [9]. For the
continuous case, let $f(X)$ be a density function associated with a
random variable $X$, where $X$ could be a real or complex scalar,
vector or matrix variable. In the present paper we consider only the
real cases for convenience. Let
$$M_{\alpha}(f)=\frac{\int_X[f(X)]^{2-\alpha}{\rm
d}X-1}{\alpha-1},~\alpha\ne 1.\eqno(1.1)
$$Note that when $\alpha\to 1, M_{\alpha}(f)\to S(f)=-\int_Xf(X)\ln
f(X){\rm d}X$ where $S(f)$ is Shannon's entropy [9] and in this
sense (1.1) is a $\alpha$-generalized entropy measure. The
corresponding discrete case is available as
$$\frac{\sum_{i=1}^kp_i^{2-\alpha}-1}{\alpha-1},p_i>0,i=1,...,k,p_1+...p_k=1,\alpha\ne
1.\nonumber$$ Characterization properties and applications of (1.1)
may be seen from [9]. Note that
$$\int_X[f(X)]^{2-\alpha}{\rm d}X=\int_X[f(X)]^{1-\alpha}f(X){\rm
d}X=E[f(X)]^{1-\alpha}.\nonumber
$$Thus there is a parallelism with Kerridge's measure of inaccuracy.
The $\alpha$-generalized Kerridge's measure of inaccuracy [9] is
given by
$$\frac{\int_xP(x)[Q(x)]^{1-\alpha}-1}{\alpha-1}=\frac{E[Q(x)]^{1-\alpha}-1}{\alpha-1},\alpha\ne
1.\eqno(1.2)
$$When $\alpha\to 1$, eq. (1.2) goes to Kerridge's measure of
inaccuracy given by
$$K(P,Q)=-\int_xP(x)\ln Q(x){\rm d}x,\eqno(1.3)
$$where $x$ is a scalar variable, $P(x)$ is the true density and
$Q(x)$ is a hypothesized or assigned density for the true density
$P(x)$. Then a measure of inaccuracy in taking $Q(x)$ for the true
density $P(x)$ is given by (1.3) and its $\alpha$-generalized form
is given by (1.2). \vskip.2cm Earlier works on Shannon's measure of
entropy, measure of directed divergence, measure of inaccuracy and
related items and applications in natural sciences may be seen from [9] and the references therein.
A measure of entropy, parallel to the one of Havrda-Charvat entropy
was introduced by Tsallis in 1988 [10, 12, 13], given by
$$T_{\alpha}(f)=\frac{\int_x[f(x)]^{\alpha}{\rm
d}x-1}{1-\alpha},\alpha\ne 1.\eqno(1.4)
$$Tsallis statistics or non-extensive statistical mechanics is
derived by optimizing (1.4) by putting restrictions in an escort
density associated with $f(x)$ of (1.4). Let
$g(x)=\frac{[f(x)]^{\alpha}}{m},m=\int_x[f(x)]^{\alpha}{\rm
d}x<\infty$. If $T_{\alpha}(f)$ is optimized over all non-negative
functional $f$, subject to the conditions that $f(x)$ is a density
and the expected value in the escort density is a given quantity,
that is $\int_xxg(x){\rm d}x=$ a given quantity, then the Euler
equation to be considered, if we optimize by using calculus of
variations, is that
$$\frac{\partial}{\partial
f}[\{f(x)\}^{\alpha}-\lambda_1f(x)+\lambda_2x\{f(x)\}^{\alpha}]=0\nonumber
$$where $\lambda_1$ and $\lambda_2$ are Lagrangian multipliers. That is,
$$\alpha[f(x)]^{\alpha-1}-\lambda_1+\lambda_2 x\alpha
[f(x)]^{\alpha-1}=0.\nonumber
$$Then
$$f(x)=c[1+\lambda_2x]^{-\frac{1}{\alpha-1}},~c=(\frac{\lambda_1}{\alpha})^{\frac{1}{\alpha-1}}.\nonumber
$$Taking $\lambda_2=a(\alpha-1)$ for $\alpha>1,a>0$  we have
Tsallis statistics as
$$f(x)=c[1+a(\alpha-1)x]^{-\frac{1}{\alpha-1}},\alpha>1,a>0.\eqno(1.5)
$$For $\alpha<1$, writing $\alpha-1=-(1-\alpha)$ the density in
(1.5) changes to
$$f_x(x)=c_1[1-a(1-\alpha)x]^{\frac{1}{1-\alpha}},\alpha<1,a>0\nonumber,
$$where $1-a(1-\alpha)x>0$ and $c_1$ can act as a normalizing constant if $f_1(x)$ is to be
taken as a statistical density. Tsallis statistics in (1.5) led
to the development of none-extensive statistical
mechanics. We will show later that (1.5) comes directly from the entropy of (1.1)
without going through any escort density. Let us optimize (1.1)
subject to the conditions that $f(x)$ is a density, $\int_xf(x){\rm
d}x=1$, and that the expected value of $x$ in $f(x)$ is a given
quantity, that is, $\int_xxf(x){\rm d}x=$ a given quantity. Then, if
we use calculus of variations, the Euler equation is of the
form
$$\frac{\partial}{\partial
f}[\{f(x)\}^{2-\alpha}-\lambda_1f(x)+\lambda_2 xf(x)]=0\nonumber,
$$where $\lambda_1$ and $\lambda_2$ are Lagrangian multipliers. Then
we have
$$f_1(x)=c_1[1-a(1-\alpha)x]^{\frac{1}{1-\alpha}},\alpha<1,a>0\eqno(1.6)
$$by taking $\frac{\lambda_2}{\lambda_1}=a(1-\alpha),a>0,\alpha<1,$
and $c_1$ is the corresponding normalizing constant to make $f_1(x)$
a statistical density. Now, for $\alpha>1$, write
$1-\alpha=-(\alpha-1)$, then directly from (1.6),
without going through any escort density, we have
$$f_2(x)=c_2[1+a(\alpha-1)x]^{-\frac{1}{\alpha-1}},\alpha>1,a>0,\eqno(1.7)
$$which is Tsallis statistics for $\alpha>1$. Thus, both the cases
$\alpha<1$ and $\alpha>1$ follow directly from (1.1). \vskip.2cm Now,
let us look into optimizing (1.1) over all non-negative integrable
functionals, $f(x)\ge 0$ for all $x$, $\int_xf(x){\rm d}x<\infty$,
such that two moment-type relations are imposed on $f$, of the form
$$\int_xx^{\gamma (1-\alpha)}f(x){\rm d}x=\mbox{ given, and }
\int_xx^{\gamma(1-\alpha)+\delta}f(x){\rm d}x=\mbox{
given.}\eqno(1.8)
$$Then the Euler equation becomes
$$\frac{\partial}{\partial f}[\{f(x)\}^{2-\alpha}-\lambda_1
x^{\gamma(1-\alpha)}f(x)+\lambda_2x^{\gamma(1-\alpha)+\delta}f(x)]=0,\nonumber
$$which leads to
$$f_1^{*}(x)=c_1^{*}x^{\gamma}[1-a(1-\alpha)x^{\delta}]^{\frac{1}{1-\alpha}},a>0,\alpha<1,\delta>0,\gamma>0\eqno(1.9)
$$for $1-a(1-\alpha)x^{\delta}>0$, by taking $\frac{\lambda_2}{\lambda_1}=a(1-\alpha),a>0,\alpha<1$,
where $c_1^{*}$ can act as the normalizing constant. Eq. (1.9) is a
special case of the pathway model of [4] for the real scalar
positive random variable $x>0$. For $\gamma=0,\delta=1$ in (1.9) we
obtain Tsallis statistics of (1.6) for the case $\alpha<1$. When
$\alpha>1$ write $1-\alpha=-(\alpha-1)$ for $\alpha>1$ then (1.9)
becomes
$$f_2^{*}(x)=c_2^{*}x^{\gamma}[1+a(\alpha-1)x^{\delta}]^{-\frac{1}{\alpha-1}},\alpha>1,a>0,x>0,\delta>0.\eqno(1.10)
$$When $\alpha\to 1$ both $f_1^{*}(x)$ of (1.9) and $f_2^{*}(x)$ of
(1.10) go to
$$f_3^{*}(x)=c_3^{*}x^{\gamma}{\rm
e}^{-ax^{\delta}},a>0,\delta>0,x>0.\eqno(1.11)
$$Eq. (1.10) for $\alpha>1, x>0$ is superstatistics [1,2].

\vskip.3cm\noindent{\bf 2.\hskip.3cm A Generalized Measure of
Entropy}

\vskip.3cm Let $X$ be a scalar, a $p\times 1$ vector of scalar
random variables or a $p\times n, p\ge n$ matrix of rank $n$ of
scalar random variables and let $f(X)$ be a real-valued scalar
function such that $f(X)\ge 0$ for all $X$ and $\int_Xf(X){\rm
d}X=1$ where ${\rm d}X$ stands for the wedge product of the
differentials in $X$. For example, if $X$ is $m\times n$,
$X=(x_{ij})$ then

$${\rm d}X=\prod_{i=1}^m\prod_{j=1}^n\wedge{\rm d}x_{ij},\nonumber
$$where $\wedge$ stands for the wedge product of differentials,
${\rm d}x\wedge{\rm d}y=-{\rm d}y\wedge{\rm d}x\Rightarrow {\rm
d}x\wedge{\rm d}x=0$. Then $f(X)$ is a density of $X$. When $X$ is
$p\times n,p\ge n$ we have a rectangular matrix variate density. For
convenience we have taken $X$ of full rank $n\le p$. When $n=1$ we
have a multivariate density and when $n=1,p=1$ we have a univariate
density. Consider the generalized entropy of (1.1) for this matrix
variate density, denoted by $f(X)$, then
$$M_{\alpha}(f)=\frac{\int_X[f(X)]^{2-\alpha}{\rm
d}X-1}{\alpha-1},\alpha\ne 1.\eqno(2.1)
$$Let $n=1$. Let us consider the situation of the ellipsoid of
concentration being a preassigned quantity. Let $X$ be $p\times 1$
vector random variable. Let $V=E[(X-E(X))(X-E(X))']>O$ (positive
definite) where $E$ denotes expected value. For convenience let us
denote $E(X)=\mu$. Then $\rho=E[(X-\mu)'V^{-1}(X-\mu)]$ is the
ellipsoid of concentration. Let us optimize (2.1) subject to the
constraint that $f(X)\ge 0$ is a density and that the ellipsoid of
concentration over all functional $f$ is a constant, that is,
$\int_Xf(X){\rm d}X=1$ and
$\int_X[(X-\mu)'V^{-1}(X-\mu)]^{\delta}f(X){\rm d}X=$ given, where
$\delta>0$ is a fixed parameter. If we are using calculus of
variation then the Euler equation is given by
$$\frac{\partial}{\partial
f}[\{f(X)\}^{2-\alpha}-\lambda_1f(X)+\lambda_2[(X-\mu)'V^{-1}(X-\mu)]^{\delta}f(X)]=0,\nonumber
$$where $\lambda_1$ and $\lambda_2$ are Lagrangian multipliers.
Solving the above equation we have
$$f_1(X)=C_1[1-a(1-\alpha)\{(X-\mu)'V^{-1}(X-\mu)\}^{\delta}]^{\frac{1}{1-\alpha}}\eqno(2.2)
$$for $\alpha<1,a>0$ where we have taken
$\frac{\lambda_2}{\lambda_1}=a(1-\alpha),a>0,\alpha<1$ and
$(\frac{\lambda_1}{2-\alpha})^{\frac{1}{1-\alpha}}=C_1$. This $C_1$
can act as the normalizing constant to make $f(X)$ in (2.2) a
statistical density. Note that for $\alpha>1$, we have from (2.2)
$$f_2(X)=C_2[1+a(\alpha-1)\{(X-\mu)'V^{-1}(X-\mu)\}^{\delta}]^{-\frac{1}{\alpha-1}},\alpha>1,a>0,\eqno(2.3)
$$and when $\alpha\to 1$, $f_1$ and $f_2$ go to
$$f_3(X)=C_3{\rm e}^{-a[(X-\mu)'V^{-1}(X-\mu)]^{\delta}}.\eqno(2.4)
$$Eq. (2.4) for $\delta =1$ is the multivariate Gaussian density.
If $Y=V^{-\frac{1}{2}}(X-\mu)$, where $V^{-\frac{1}{2}}$ is the
positive definite square root of the positive definite matrix
$V^{-1}$, then ${\rm d}Y=|V|^{-\frac{1}{2}}{\rm d}X$ and the density
of $Y$, denoted by $g(Y)$, is given by
$$g(Y)=C_4~{\rm
e}^{-a(y_1^2+...+y_p^2)^{\delta}},-\infty<y_j<\infty,j=1,...,p,Y'=(y_1,...,y_p)\eqno(2.5)
$$and $C_4$ is the normalizing constant. This normalizing constant
can be evaluated in two different ways. One method is to use polar
coordinate transformation, see Theorem 1.25 of [3]. Let
\begin{align}
y_1&=r~\sin\theta_1\sin\theta_2...\sin\theta_{p-1}\nonumber\\
y_2&=r~\sin\theta_1...\sin\theta_{p-2}\cos\theta_{p-1}\nonumber\\
\vdots&=\vdots\nonumber\\
y_{p-1}&=r~\sin\theta_1\cos\theta_1\nonumber\\
y_p&=r~\cos\theta_1,\nonumber
\end{align}
where $r>0,0<\theta_j\le \pi,j=1,...,p-2,0<\theta_{p-1}\le 2\pi$
and the Jacobian is given by
$${\rm d}y_1\wedge...\wedge{\rm
d}y_p=r^{p-1}\{\prod_{j=1}^{p-1}|\sin\theta_j|^{p-j-1}\}{\rm
d}r\wedge{\rm d}\theta_1\wedge...\wedge{\rm
d}\theta_{p-1}.\eqno(2.6)
$$Under this transformation the exponent
$(y_1^2+...+y_p^2)^{\delta}=(r^2)^{\delta}$. Hence we integrate out
the sine functions. The integral over $\theta_{p-1}$ goes from $0$
to $2\pi$ and gives the value $2\pi$, and others from $0$ to $\pi$.
These, in general, can be evaluated by using type-1 beta integrals
by putting $\sin\theta =u$ and $u^2=v$. That is,
\begin{align}
\int_0^{\pi}\sin\theta~{\rm d}\theta&=2\int_0^{\pi/2}\sin\theta~{\rm
d}\theta=2\int_0^1u(1-u^2)^{-\frac{1}{2}}{\rm d}u\nonumber\\
&=\int_0^1v^{1-1}(1-v)^{-\frac{1}{2}}{\rm
d}v=\frac{\Gamma(1)\Gamma(1/2)}{\Gamma(3/2)}\nonumber\\
\int_0^{\pi}(\sin\theta)^2{\rm
d}\theta&=\frac{\Gamma(3/2)\Gamma(1/2)}{\Gamma(4/2)}\nonumber\\
\vdots&=\vdots\nonumber\\
\int_0^{\pi}(\sin\theta)^{p-2}{\rm
d}\theta&=\frac{\Gamma(\frac{p-1}{2})\Gamma(1/2)}{\Gamma(\frac{p}{2})}.\nonumber
\end{align}
Taking the product we have
$$2\pi\frac{(\sqrt{\pi})^{p-2}}{\Gamma(\frac{p}{2})}=\frac{2\pi^{p/2}}{\Gamma(p/2)}.\nonumber
$$Hence the total integral is equal to
$$1=C_4|V|^{\frac{1}{2}}\frac{2\pi^{p/2}}{\Gamma(p/2)}\int_0^{\infty}r^{p-1}{\rm
e}^{-ar^{2\delta}}{\rm d}r,\delta>0.\nonumber
$$Put $x=ar^{2\delta}$ and integrate out by using a gamma integral
to get
$$C_4=\frac{\delta\Gamma(\frac{p}{2})a^{\frac{p}{2\delta}}}{|V|^{\frac{1}{2}}\pi^{p/2}\Gamma(\frac{p}{2\sigma})}.$$
That is, the density is given by
$$f_3(X)=\frac{\delta~a^{\frac{p}{2\delta}}\Gamma(p/2)}{|V|^{1/2}\pi^{p/2}\Gamma(\frac{p}{2\delta})}{\rm
e}^{-a[(X-\mu)'V^{-1}(X-\mu)]^{\delta}},\delta>0,a>0,V>O.\eqno(2.7)
$$From the above steps the following items are available: The
density of $Y=V^{-\frac{1}{2}}(X-\mu)$ is available as
$$g(Y)=\frac{\delta~a^{\frac{p}{2\delta}}\Gamma(\frac{p}{2})}{\pi^{p/2}\Gamma(\frac{p}{2\delta})}{\rm
e}^{-a(Y'Y)^{\delta}}.\eqno(2.8)
$$The density of $u=Y'Y=y_1^2+...+y_p^2$, denoted by $g_1(u)$, is
given by
$$g_1(u)=\frac{\delta~a^{\frac{p}{2\delta}}}{\Gamma(\frac{p}{2\delta})}u^{\frac{p}{2}-1}{\rm
e}^{-au^{\delta}},\delta>0,u>0,\eqno(2.9)
$$and the density of $r>0$, where $r^2=u=Y'Y$, denoted by $g_2(r)$,
is given by
$$g_2(r)=\frac{2\delta~a^{\frac{p}{2\delta}}}{\Gamma(\frac{p}{2\delta})}r^{p-1}{\rm
e}^{-ar^{2\delta}},r>0,\delta>0.\eqno(2.10)
$$
{\bf 2.1. Another method}\hskip.3cm Another direct way of deriving the
densities of $X,Y=V^{-\frac{1}{2}}(X-\mu),u=Y'Y,r=\sqrt{u}$ is the
following: From [3] see the transformation in Stiefel manifold where
a matrix of the form $n\times p, n\ge p$ of rank $p$ is transformed
into $S=X'X$ which is a $p\times p$ matrix, where the differential
elements, after integrating out over the Stiefel manifold, are
connected by the relation, see also Theorem 2.16 and Remark 2.13 of
[3],
$${\rm
d}X=\frac{\pi^{\frac{np}{2}}}{\Gamma_p(\frac{n}{2})}|S|^{\frac{n}{2}-\frac{p+1}{2}}{\rm
d}S\eqno(2.11)
$$where $|S|$ denotes the determinant of $S$ and $\Gamma_p(\alpha)$
is the real matrix-variate gamma given by
$$\Gamma_p(\alpha)=\pi^{\frac{p(p-1)}{4}}\Gamma(\alpha)\Gamma(\alpha-\frac{1}{2})...\Gamma(\alpha-\frac{p-1}{2}),
\Re(\alpha)>\frac{p-1}{2}.\eqno(2.12)
$$Applications of the above result in various disciplines may be seen from [5,6,7,8]. In our problem, we can connect ${\rm d}Y$ of (2.8) to ${\rm d}u$
of (2.9) with the help of (2.11) by replacing $n$ by $p$ and $p$ by
$1$ in the $n\times p$ matrix. That is, from (2.11)
$${\rm d}Y=\frac{\pi^{p/2}}{\Gamma(p/2)}u^{\frac{p}{2}-1}{\rm
d}u.\eqno(2.13)
$$The total integral of $f_3(X)$ of (2.3) is given by
$$1=\int_Xf_3(X){\rm
d}X=C_3|V|^{1/2}\frac{\pi^{p/2}}{\Gamma(p/2)}\int_{u=0}^{\infty}u^{\frac{p}{2}-1}{\rm
e}^{-au^{\delta}}{\rm d}u,a>0,\delta>0.\nonumber
$$Put $v=au^{\delta}$ and integrate out by using a gamma integral to
get
$$C_3=\frac{\delta~a^{\frac{p}{2\delta}}\Gamma(p/2)}{|V|^{1/2}\pi^{p/2}\Gamma(\frac{p}{2\delta})}\nonumber
$$and we get the same result as in (2.7), thereby the same
expressions for $g(Y)$ in (2.8), $g_1(u)$ in (2.9) and $g_2(r)$ in
(2.10).

\vskip.3cm\noindent{\bf 3.\hskip.3cm A Generalized Model}

\vskip.3cm If we optimize (2.1) over all integrable functions
$f(X)\ge 0$ for all $X$, subject to the two moment-like restrictions
$E[(X-\mu)'V^{-1}(X-\mu)]^{\gamma(1-\alpha)}=$ fixed and
$E[(X-\mu)'V^{-1}(X-\mu)]^{\delta+\gamma(1-\alpha)}=$ fixed, then the
corresponding Euler equation becomes
$$\frac{\partial}{\partial
f}[\{f(X)\}^{2-\alpha}-\lambda_1[(X-\mu)'V^{-1}(X-\mu)]^{\gamma(1-\alpha)}+\lambda_2[(X-\mu)V^{-1}(X-\mu)]^{\delta+\gamma(1-\alpha)}]=0\nonumber
$$and the solution is available as
$$f(X)=C^{*}[(X-\mu)'V^{-1}(X-\mu)]^{\gamma}[1-a(1-\alpha)\{(X-\mu)'V^{-1}(X-\mu)\}^{\delta}]^{\frac{1}{1-\alpha}}\eqno(3.1)
$$for $\alpha<1,a>0, V>O, \delta>0,\gamma>0$ and for convenience we
have taken $\frac{\lambda_2}{\lambda_1}=a(1-\alpha),a>0,\alpha<1$,
where $C^{*}$ can act as the normalizing constant if $f(X)$ is to be
treated as a statistical density. Otherwise $f(X)$ can be a very
versatile model in model building situations. If $C^{*}$ is the
normalizing constant then it can be evaluated by using the following
procedure: Put $Y=V^{-\frac{1}{2}}(X-\mu)\Rightarrow {\rm
d}Y=|V|^{-\frac{1}{2}}{\rm d}X$. The total integral is 1, that is,
$$1=\int_Xf(X){\rm
d}X=C^{*}|V|^{\frac{1}{2}}\int_Y[Y'Y]^{\gamma}[1-a(1-\alpha)(Y'Y)^{\delta}]^{\frac{1}{1-\alpha}}{\rm
d}Y.\nonumber
$$Let $u=Y'Y$, then ${\rm
d}Y=\frac{\pi^{p/2}}{\Gamma(p/2)}u^{\frac{p}{2}-1}{\rm d}u$ from
(2.13). Then for $a>0,\alpha<1,\delta>0$ we can integrate out by
using a type-1 beta integral by putting $z=a(1-\alpha)u^{\delta}$
for $\alpha<1$. Then the normalizing constant, denoted by $C_1^{*}$,
is available as
$$C_1^{*}=\frac{\delta[a(1-\alpha)]^{\frac{\gamma}{\delta}+\frac{p}{2\delta}}
\Gamma(p/2)\Gamma(\frac{1}{1-\alpha}+1+\frac{\gamma}{\delta}+\frac{p}{2\delta})}{|V|^{1/2}\pi^{p/2}
\Gamma(\frac{\gamma}{\delta}+\frac{p}{2\delta})\Gamma(1+\frac{1}{1-\alpha})},\eqno(3.2)
$$for $\delta>0,~\gamma+\frac{p}{2}>0$. Hence the density of the $p\times 1$ vector $X$ is given by

$$f_1(X)=C_1^{*}[(X-\mu)'V^{-1}(X-\mu)]^{\gamma}[1-a(1-\alpha)[(X-\mu)'V^{-1}(X-\mu)]^{\delta}]^{\frac{1}{1-\alpha}}
\eqno(3.3)
$$for $V>O,a>0,\delta>0,\gamma+\frac{p}{2}>0$,
$X'=(x_1,...,x_p),\mu'=(\mu_1,...,\mu_p)$,
$-\infty<x_j<\infty,-\infty<\mu_j<\infty,j=1,...,p$. For $\alpha<1$
we may say that $f(X)$ in (3.3) is a generalized type-1 beta form.
Then the density of $Y$, denoted by $g(Y)$, is given by
$$g(Y)=|V|^{1/2}C_1^{*}(Y'Y)^{\gamma}[1-a(1-\alpha)(Y'Y)^{\delta}]^{\frac{1}{1-\alpha}},\nonumber
$$for $a>0,\alpha<1$ and $C_1^{*}$ is defined in (3.2). Note that
the density of $u=Y'Y$, denoted by $g_1(u)$, is available, as
$$g_1(u)=\tilde{C_1}u^{\gamma+\frac{p}{2}-1}[1-a(1-\alpha)u^{\delta}]^{\frac{1}{1-\alpha}},\eqno(3.4)
$$where
$$\tilde{C_1}=\frac{\delta[a(1-\alpha)]^{\frac{\gamma}{\delta}+\frac{p}{2\delta}}\Gamma(\frac{1}{1-\alpha}+1+\frac{\gamma}{\delta}+\frac{p}{2\delta})}
{\Gamma(\frac{\gamma}{\delta}+\frac{p}{2\delta})\Gamma(\frac{1}{1-\alpha}+1)},
$$for $\delta>0,~ \gamma+\frac{p}{2}>0$. Note that for $\alpha>1$ in (3.1) the model switches into a
generalized type-2 beta form. Write $1-\alpha=-(\alpha-1)$ for
$\alpha>1$. Then the model in (3.2) switches into the following
form:
$$f_2(X)=C_2^{*}[(X-\mu)'V^{-1}(X-\mu)]^{\gamma}[1+a(\alpha-1)[(X-\mu)'V^{-1}(X-\mu)]^{\delta}]^{-\frac{1}{\alpha-1}}
\eqno(3.5)
$$for $\delta>0,a>0,V>O,\alpha>1$. The normalizing constant
$C_2^{*}$ can be computed by using the following procedure. Put
$z=a(\alpha-1)u^{\delta},\delta>0,\alpha>1$. Then integrate out by
using a type-2 beta integral to get
$$C_2^{*}=\frac{\delta[a(\alpha-1)]^{\frac{\gamma}{\delta}+\frac{p}{2\delta}}\Gamma(p/2)\Gamma(\frac{1}{\alpha-1})}{|V|^{1/2}\pi^{p/2}
\Gamma(\frac{\gamma}{\delta}+\frac{p}{2\delta})\Gamma(\frac{1}{\alpha-1}-\frac{\gamma}{\delta}-\frac{p}{2\delta})}\eqno(3.6)
$$for $\gamma+p/2>0$, $\frac{1}{\alpha-1}-\frac{\gamma}{\delta}-\frac{p}{2\delta}>0$. When $\alpha\to 1$ then both $f_1(X)$ of (3.3) and $f_2(X)$ of
(3.5) go to the generalized gamma model given by
$$f_3(X)=C_3^{*}[(X-\mu)'V^{-1}(X-\mu)]^{\gamma}{\rm
e}^{-a[(X-\mu)'V^{-1}(X-\mu)]^{\delta}}\eqno(3.7)
$$where
$$C_3^{*}=\frac{\delta\Gamma(p/2)a^{\frac{\gamma}{\delta}+\frac{p}{2\delta}}}{|V|^{1/2}\pi^{p/2}
\Gamma(\frac{\gamma}{\delta}+\frac{p}{2\delta})},~\delta>0,~\gamma+\frac{p}{2}>0.\eqno(3.8)
$$It is not difficult to show that when $\alpha\to 1$ both
$C_1^{*}\to C_3^{*}$ and $C_2^{*}\to C_3^{*}$. This can be seen by
using Stirling's formula
$$\Gamma(z+\eta)\approx \sqrt{2\pi}z^{z+\eta-\frac{1}{2}}{\rm
e}^{-z}
$$for$|z|\to\infty$ and $\eta$ is a bounded quantity. Observe that
$$\lim_{\alpha\to 1_{-}}\frac{1}{1-\alpha}=\infty\mbox{  and
}\lim_{\alpha\to 1_{+}}\frac{1}{\alpha-1}=\infty
$$and we can apply Stirlling's formula by taking
$z=\frac{1}{1-\alpha}$ in one case and $z=\frac{1}{\alpha-1}$ in the
other case. Thus, from $f_1(X)$ we can switch to $f_2(X)$ to
$f_3(X)$ or through the same model we can go to three different
families of functions through the parameter $\alpha$ and hence
$\alpha$ is called the pathway parameter and the model above belongs
to the pathway model in [4].

\vskip.3cm\noindent{\bf 4.\hskip.3cm Generalization to the Matrix Case}

\vskip.3cm Let $X$ be a $p\times n,n\ge p$ rectangular matrix of
full rank $p$. Let $A>O$ be $p\times p$ and $B>O $ be $n\times n$
positive definite constant matrices. Let $A^{1/2}$ and $B^{1/2}$
denote the positive definite square roots of $A$ and $B$
respectively. Consider the matrix
$$I-a(1-\alpha)A^{1/2}XBX'A^{1/2}>O,\nonumber
$$where $a>0,\alpha<1.$ Let $f(X)$ be a real-valued function of $X$
such that $f(X)\ge 0$ for all $X$ and $f(X)$ is integrable,
$\int_Xf(X){\rm d}X<\infty$. If we assume that the expected value of
the determinant of the above matrix is fixed over all functional
$f$, that is
$$E|I-a(1-\alpha)A^{1/2}XBX'A^{1/2}|=\mbox{fixed},\eqno(4.1)
$$then, if we optimize the entropy (2.1) under the restriction (4.1)
the Euler equation is,
$$\frac{\partial}{\partial
f}[\{f(X)\}^{2-\alpha}-\lambda|I-a(1-\alpha)A^{1/2}XBX'A^{1/2}|f(X)]=0.\nonumber
$$Equation such as the one in (4.1) can be connected to the volume of a certain parallelotope or random geometrical objects. Solving it we have
$$f(X)=\hat{C}|I-a(1-\alpha)A^{1/2}XBX'A^{1/2}|^{\frac{1}{1-\alpha}}\eqno(4.2)
$$where $\hat{C}$ is a constant. A more general form is to put a
restriction of the form that the expected value of
$|A^{1/2}XBX'A^{1/2}|^{\gamma(1-\alpha)}|I-a(1-\alpha)A^{1/2}XBX'A^{1/2}|$
is a fixed quantity over all functional $f$. Then
$$f(X)=\hat{C_1}|A^{1/2}XBA^{1/2}|^{\gamma}|I-a(1-\alpha)A^{1/2}XBA^{1/2}|^{\frac{1}{1-\alpha}}\eqno(4.3)
$$for $\alpha<1,a>0,A>O,B>O$ and $X$ is $p\times n,n\ge p$ of full
rank $p$ and a prime denotes the transpose. The model in (4.3) can
switch around to three functional forms, one family for $\alpha<1$,
a second family for $\alpha>1$ and a third family for $\alpha\to 1$.
In fact (4.3) contains all matrix variate statistical densities in
current use in physical and engineering sciences. For evaluating the
normalizing constants for all the three cases, the first step is to
make the transformation
$$Y=A^{1/2}XB^{1/2}\Rightarrow {\rm d}Y=|A|^{n/2}|B|^{p/2}{\rm
d}X,\eqno(4.4)
$$see [3] for the Jacobian of this transformation. After this stage,
all the steps in the previous sections are applicable and we use
matrix variate type-1 beta, type-2 beta, and gamma integrals to do
the final evaluation of the normalizing constants. Since the steps
are parallel the details are omitted here.

\vskip.3cm\noindent{\bf 5.\hskip.3cm Standard Deviation and Diffusion Entropy Analysis}

\vskip.3cm Scale invariance has been found to hold for complex systems and the correct evaluation of the scaling exponents is of fundamental importance to assess if universality classes exist. Diffusion is typically quantified in terms of a relationship between fluctuation of a variable $x$ and time $t$. A widely used method of analysis of complexity rests on the assessment of the scaling exponent of the diffusion process generated by a time series. According to the prescription of Peng et al. [14], the numbers of a time series are interpreted as generating diffusion fluctuations and one shifts the attention from the time series to the probability density function (pdf) $p(x,t)$, where $x$ denotes the variable collecting the fluctuations and $t$ is the diffusion time. In this case, if the time series is stationary, the scaling property of the pdf of the diffusion process takes the form
$$p(x,t) = \frac{1}{t^\delta}~F\left( \frac{x}{t^\delta }\right),\eqno(5.1)
$$where $\delta$ is a scaling exponent. Diffusion may scale linearly with time, leading to ordinary diffusion, or it may scale nonlinearly with time, leading to anomalous diffusion. Anomalous diffusion processes can be classified as Gaussian or L\'{e}vy, depending on wether the central limit theorem (CLT) holds. CLT entails ordinary statistical mechanics. That is, it entails a Gaussian form for F in (51.) composing a random walk without temporal correlations (i.e. $\delta=0$). Due to the CLT. the probability distribution function $p(x,t)$ describing the probabilities of $x(t)$ has a finite second moment $<x^2>$, and when the second moment diverges, $x(t)$ no longer falls under the CLT and instaed indicated that the generalized central limit theorem applies. Failures of CLT mean that instead of statistical mechanics, nonextensive statistical mechanics may be utilized [12, 13]. 

Scafetta and Grigolini [15] established that Diffusion Entropy Analysis (DEA), a method of statistical analysis based on the Shannon entropy (see eq. (1.1)) of the diffusion process, determines the correct scaling exponent $\delta$even when the statistical properties, as well as the dynamic properties, are anomalous. The other methods usually adopted to detect scaling, for example the Standard Deviation Analysis (SDA), are based on the numerical evaluation of the variance. Consequently, these methods detect a power index, denoted $H$ by Mandelbrot [16] in honor of Hurst, which might depart from the scaling $\delta$ of eq. (5.1). These variance methods (cf. Fourier analysis and wavelet analysis; see [17, 18]) produce correct results in the Gaussian case, where $H=\delta$, but fail to detect the correct scaling of the pdf, for example, in the case of L\'{e}vy flight, where the variance diverges, or in the case of L\'{e}vy walk, where $\delta$ and $H$ do not coincide, being related by $\delta =1/(3-2H)$. The case $H=\delta=0.5$ is that of a completely uncorrelated random process. The case $\delta=1$ is that of a completely regular process undergoing ballistic motion. Figs. 1 to 4 clearly show that the diffusion entropy development over time for solar neutrinos does neither meet the first nor the latter case. 
The Shannon entropy, eq. (1.1) for the diffusion process at time $t$, is defined by
$$S(t) = - \int p(x,t) ~\ln [p(x,t)] ~dx.\eqno(5.2)
$$If the scaling condition of eq. (5.1) holds true, it is easy to prove that
$$S(t) = A + \delta~ \ln(t),\eqno(5.3)
$$where
$$A \equiv -\int_{-\infty}^{\infty} dy \, F(y) \, \ln [F(y)],\eqno(5.4)
$$and $y = x/t^{\delta}$. Numerically, the scaling coefficient $\delta$ can be evaluated by using fitting curves with the form (5.3) that on a linear-log scale is a straight line. Even though time series extracted from complex environments may not show a pure scaling behaviour as in eq. (5.3) but, instead, patterns with oscillations due to periodicities, one can still observe how diffusion entropy grows linearly with time and one can estimate the diffusion exponent with reasonable accuracy. 

Figs. 1 and 2 and Figs. 3 and 4, respectively, are showing diffusion entropy as a function of time for two different time series. Figs. 1 to 4 show the numerical results of Standard Deviation Analysis and Diffusion Entropy Analysis for solar neutrino data taken by the SuperKamiokande experiments I (SK-I, 1996-2001, 1496 days, 5.0-20.0 MeV) and II (SK-II, 2002-2005, 791 days, 8.0-20.0 MeV). SuperKamiokande [16] is a 50 kiloton water Cherenkov detector located at the Kamioka Observatory of the Institute for Cosmic Ray Research, University of Tokyo. It was designed to study solar neutrino oscillations and carry out searches for the decay of the nucleon. The SuperKamiokande experiment began in 1996 and in the ensuing decade of running has produced extremely important results in the fields of atmospheric and solar neutrino oscillations, along with setting stringent limits on the decay of the nucleon and the existence of dark matter and astrophysical sources of neutrinos. Perhaps most crucially, Super-Kamiokande for the first time definitely showed that neutrinos have mass and undergo flavor oscillations.

An additional feature of the $S(t)$ behavior over time in Figs. 2 and 4 are distinct oscillations characteristic for processes with periodic modulation and asymptotic saturation. They appear for large $\delta$. At the current stage of research the origin of these oscillations is not clear.    

\begin{center}
\resizebox{12cm}{!}{\includegraphics{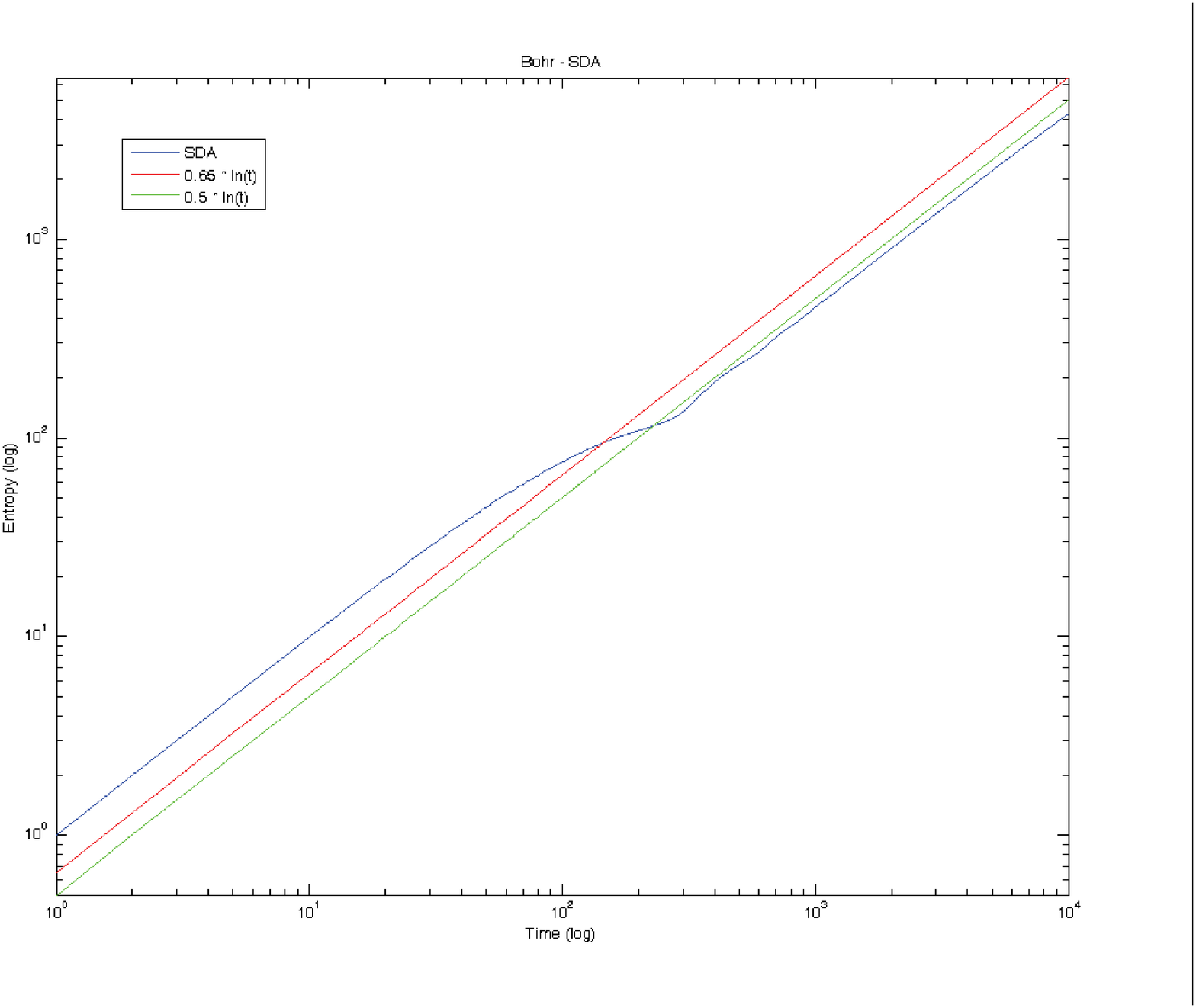}}
\end{center}
\noindent
{\small
{\bf Figure 1:} Standard Diffusion Analysis of the boron solar neutrino data from SuperKamiokande I and II. The green line coincides with a straight line with the slope $\delta = 0.5$. The red line reflects the approximated straight slope of the real data with $\delta = 0.65$. The exact result of the SDA is shown by the blue line and indicates a change in the diffusion entropy over time from $\delta > 0.5$ to $\delta = 0.5$.}\\

\begin{center}
\resizebox{12cm}{!}{\includegraphics{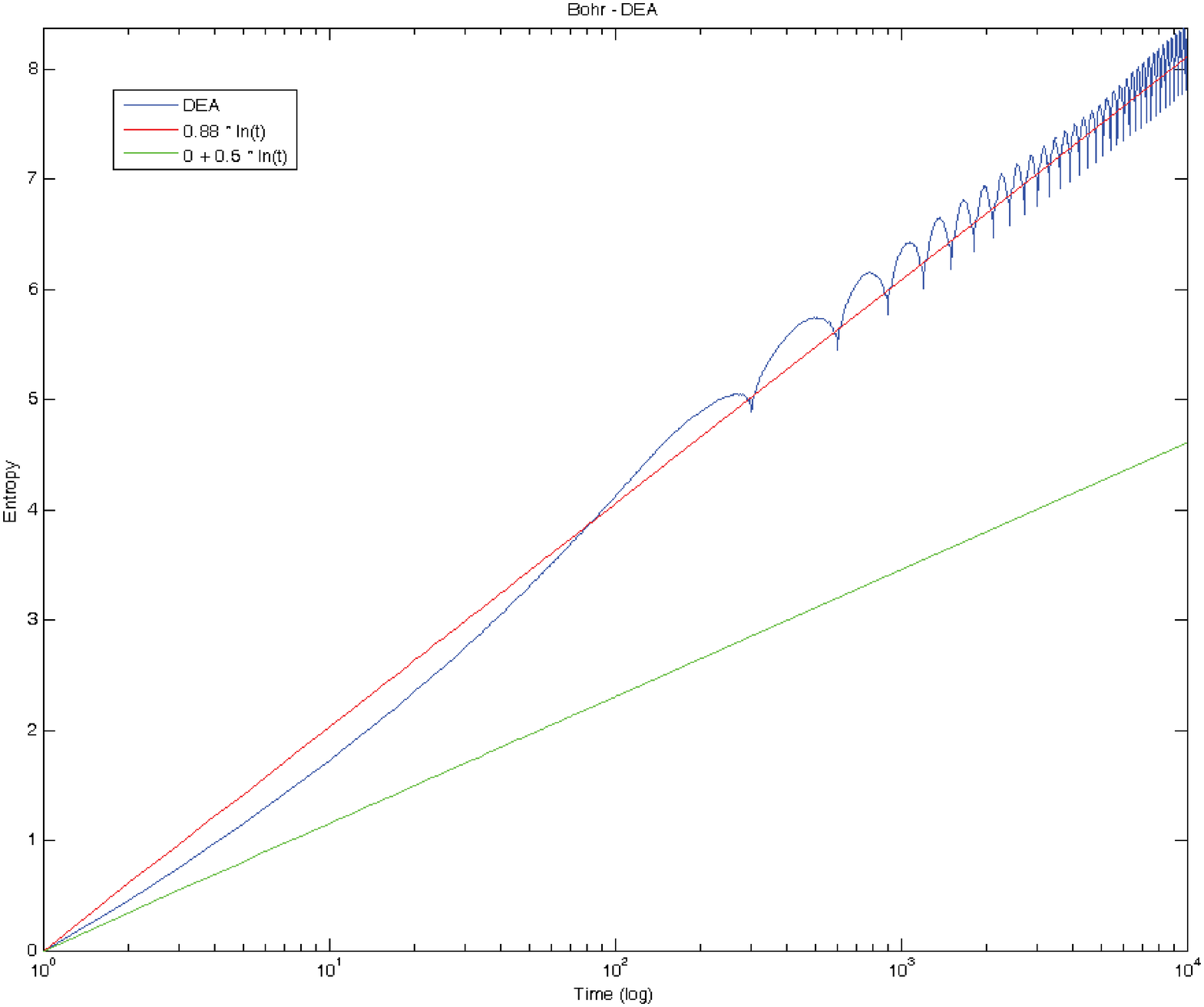}}
\end{center}
\noindent
{\small
{\bf Figure 2:} Diffusion Entropy Analysis of the boron solar neutrino data from SuperKamiokande I and II. The green line coincides with a straight line with the slope $\delta = 0.5$. The red line reflects the approximated straight slope of the real data with $\delta = 0.88$. In comparison with Fig. 1, the green and red lines are remarkable different from each other and indicate strong anomalous diffusion. The exact result of the DEA is shown by the blue line and indicates a development over time from periodic modulation to asymptotic saturation.}\\

\begin{center}
\resizebox{12cm}{!}{\includegraphics{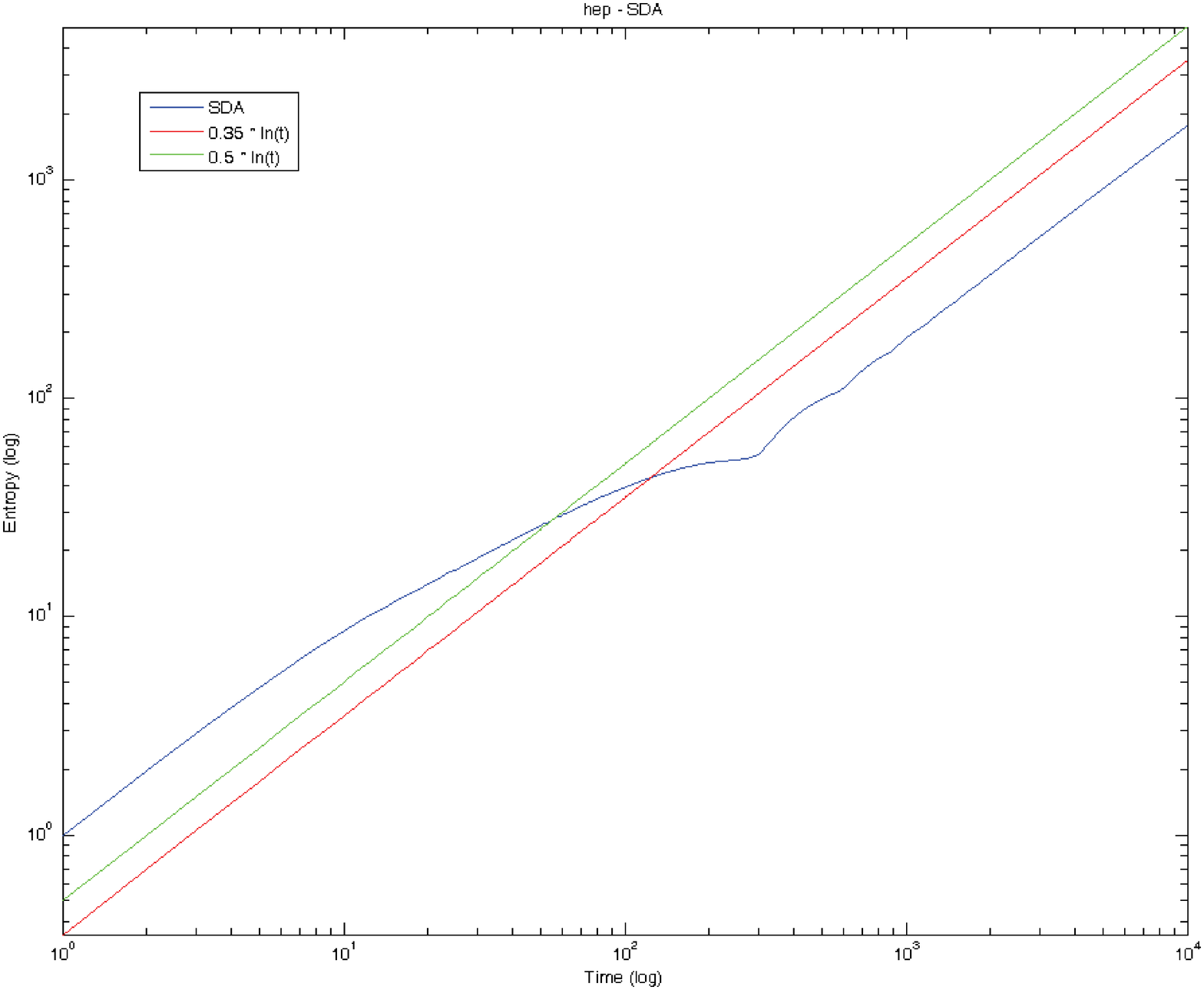}}
\end{center}
\noindent
{\small
{\bf Figure 3:} Standard Diffusion Analysis of the hep solar neutrino data from SuperKamiokande I and II. The green line coincides with a straight line with the slope $\delta = 0.5$. The red line reflects the approximated straight slope of the real data with $\delta = 0.35$. Note the remarkable difference between the boron analysis results $\delta > 0.5$ and the hep analysis results shown in this Fig. with $\delta < 0.5$. This is an indication of superdiffusion in the first case and subdiffusion in the second case. The exact result of the SDA is shown by the blue line and indicates a change in the diffusion entropy over time from $\delta > 0.5$ to $\delta < 0.5$.}\\

\begin{center}
\resizebox{12cm}{!}{\includegraphics{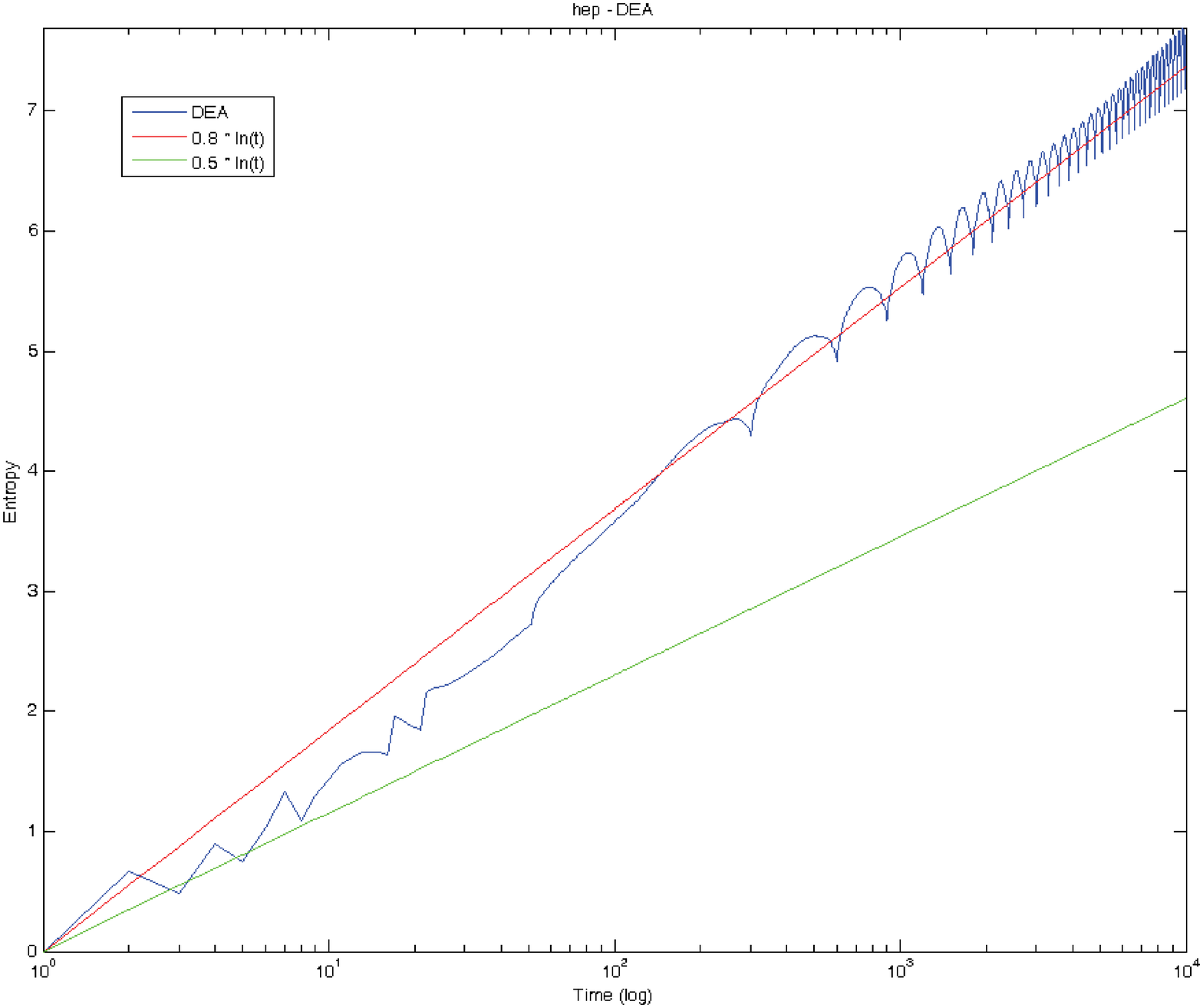}}
\end{center}
\noindent
{\small
{\bf Figure 4:} Diffusion Entropy Analysis of the hep solar neutrino data from SuperKamiokande I and II. The green line coincides with a straight line with the slope $\delta = 0.5$. The red line reflects the approximated straight slope of the real data with $\delta = 0.8$. In comparison with Fig. 3, the green and red lines are remarkable different from each other similar to the boron data analysis and indicate strong anomalous diffusion. The exact result of the DEA is shown by the blue line and indicates a development over time from periodic modulation to asymptotic saturation similar to the boron analysis results.}\\

 \vskip.3cm\noindent{\bf Acknowledgement}

 \vskip.3cm The authors would like to thank the Department of Science and Technology, Government of India, for financial assistance for this work under project No.SR/S4/MS:287/05. The authors are also grateful to Dr. Alexander Haubold, Columbia University New York, for the numerical analysis of the solar neutrino data with Standard Deviation Analysis and Diffusion Entropy Analysis.

 \vskip.3cm\centerline{\bf References}

 \vskip.2cm\noindent[1]~~C. Beck, Stretched exponentials from superstatistics, Physica A, 365(2006), 96-101.

 \vskip.2cm\noindent [2]~~C. Beck and E.G.D. Cohen, Superstatistics, Physica A, 322(2003), 267-275.

 \vskip.2cm\noindent[3]~~A.M. Mathai, Jacobians of Matrix Transformations and Functions of Matrix Argument, World Scientific Publishing,
 New York, 1997.

 \vskip.2cm\noindent [4]~~A.M. Mathai, A pathway to matrix variate gamma and normal densities, Linear Algebra and its
 Applications, 396(2005), 317-328.

 \vskip.2cm\noindent[5]~~A.M. Mathai, Some properties of Mittag-Leffler functions and matrix-variate analogues: A
 statistical perspective, Fractional Calculus \& Applied Analysis, 13(1)(2010), 113-132.

 \vskip.2cm\noindent [6]~~A.M. Mathai and H.J. Haubold, Pathway model, superstatistics, Tsallis statistics and a generalized
 measure of entropy, Physica A, 375(2007), 110-122.

 \vskip.2cm\noindent[7]~~A.M. Mathai and H.J. Haubold, Special Functions for Applied Scientists, Springer, New York, 2008.

 \vskip.2cm\noindent[8]~~A.M. Mathai, S.B. Provost, and T. Hayakawa, Bilinear Forms and Zonal Polynomials, Springer, New York, 1995.

 \vskip.2cm\noindent [9]~~A.M. Mathai and P.N. Rathie, Basic Concepts in Information Theory and Statistics: Axiomatic
 Foundations and Applications, Wiley Eastern, New Delhi and Wiley Halsted, New York, 1975.

 \vskip.2cm\noindent [10]~~C. Tsallis, Possible generalizations of Boltzmann-Gibbs statistics, Journal of Statistical Physics, 52(1988), 479-487.
 
 \vskip.2cm\noindent [11]~~A. Greven, G. Keller, and G. Warnecke (Eds.), Entropy, Princeton University Press, Princeton and Oxford, 2003.

 \vskip.2cm\noindent [12]~~C. Tsallis, Introduction to Nonextensive Statistical Mechanics: Approaching a Complex World, Springer, New York, 2009.
 
 \vskip.2cm\noindent [13]~~M. Gell-Mann and C. Tsallis (Eds.), Nonextensive Entropy: Interdisciplinary Applications, Oxford University Press, New York, 2004.
 
 \vskip.2cm\noindent [14]~~C.K. Peng, S.V. Buldyrev, S. Havlin, M. Simons, H.E. Stanley, and A.L. Goldberger, Mosaic organization of DNA nucleotides, Physical Review E, 49(1995), 1685-1689.
 
 \vskip.2cm\noindent [15]~~N. Scafetta and P. Grigolini, Scaling detection in time series:Diffusion entropy analysis, Physical Review E, 66(2002), doi: 101103/Phys.RevE.66.036130.
 
 \vskip.2cm\noindent [15]~~B.B. Mandelbrot, The Fractal Geometry of Nature, W.H. Freeman and Company, New York, 1983.
 
 \vskip.2cm\noindent [16]~~http://www-sk.icrr.u-tokyo.ac.jp/sk/index-e.html.
 
 \vskip.2cm\noindent [17]~~H.J. Haubold and A.M. Mathai, A heuristic remark on the periodic variation in the number of solar neutrinos detected on Earth, Astrophysics and Space Science, 228(1995), 113-134.
 
 \vskip.2cm\noindent [18]~~K. Sakurai, H.J. Haubold, and T. Shirai, The variation of the solar neutrino fluxes over time in the Homestake, GALLEX(GNO), and the Super-Kamiokande experiments, Space Radiation, 5(2008), 207-216.

\end{document}